# Local chemical bonding and structural properties in $Ti_3AlC_2$ MAX phase and $Ti_3C_2T_x$ MXene probed by Ti *1s* X-ray absorption spectroscopy

Martin Magnuson and Lars-Åke Näslund

*Department of Physics, Chemistry and Biology (IFM), Linköping University, SE-581 83 Linköping, Sweden*

2020-09-30

## Abstract

The chemical bonding within the transition-metal carbide materials MAX phase $Ti_3AlC_2$ and MXene $Ti_3C_2T_x$ is investigated by X-ray absorption near-edge structure (XANES) and extended X-ray absorption fine structure (EXAFS) spectroscopies. MAX phases are inherently nanolaminated materials that consist of alternating layers of $M_{n+1}X_n$ and monolayers of an A-element from the IIIA or IVA group in the periodic table, where M is a transition metal and X is either carbon or nitrogen. Replacing the A-element with surface termination species $T_x$ will separate the $M_{n+1}X_n$-layers forming two-dimensional (2D) flakes of $M_{n+1}X_nT_x$. For $Ti_3C_2T_x$ the $T_x$ corresponds to fluorine (F) and oxygen (O) covering both sides of every single 2D $M_{n+1}X_n$-flake. The Ti *K*-edge (*1s*) XANES of both $Ti_3AlC_2$ and $Ti_3C_2T_x$ exhibit characteristic pre-edge absorption regions of C *2p* - Ti *3d* hybridization with clear crystal-field splitting's while the main-edge absorption features originate from the Ti *1s → 4p* excitation, where only the latter shows sensitivity towards the fcc-site occupation of the termination species. The coordination numbers obtained from EXAFS show that $Ti_3AlC_2$ and $Ti_3C_2T_x$ are highly anisotropic with a strong in-plane contribution for Ti and with a dynamic out-of-plane contribution from the Al monolayers and termination species, respectively. As shown in the temperature-dependent measurements, the O contribution shifts to shorter bond length while the F diminishes as the temperature is raised from room temperature up to 750 °C.

## 1. Introduction

Despite the large interest in graphene [1], which lacks a natural band gap, it has been difficult to artificially produce graphene-based materials with suitable band gaps. This has encouraged researchers to explore other two-dimensional (2D) materials such as hexagonal boron nitride (h-BN), molybdenum disulphide ($MoS_2$), tungsten disulphide ($WS_2$), and MXenes ($M_{n+1}X_nT_x$). The last example is also the latest 2D material, developed in the last decade, and consists of a family of 2D transition metal carbides denoted $M_{n+1}X_nT_x$ (n = 1, 2, 3) where M is a transition metal, X is either carbon or nitrogen, and $T_x$ denotes surface termination species [2-4]. The layered structures contain more than one element and, thus, offer properties that may be useful for transistors and spintronics [5], 2D-based electronics and screens [6] in addition to energy storage systems such as supercapacitors [7], Li-ion batteries [8], fuel- and solar cells [9] as well as transparent conductive electrodes [10] and composite materials with high strength [11].

The parent precursor compounds of MXenes are inherently nanolaminated materials known as MAX phases ($M_{n+1}AX_n$, n = 1, 2, 3) [12] (space group $P6_3/mmc$), where A is a *p*-element that usually belongs to groups IIIA or IVA in the periodic table. These phases contain more than 150 variants [13], including $Ti_3AlC_2$ that is a precursor for the $Ti_3C_2T_x$ MXene. To make MXene from $M_{n+1}AX_n$, the weakly bounded A-layers are etched away and replaced by surface termination groups ($T_x$) in the exfoliation process [2]. The delamination results in weakly bounded stacks of 2D sheets with $M_{n+1}X_nT_x$ composition.

Generally, MXenes consist of a core of a few atoms thick 2D $M_{n+1}X_n$ conductive carbide layer that is crystalline in the basal plane and a transition metal surface that can be functionalized for different material properties by changing the chemistry of the termination species. Layered structures like MXenes contains more than one element and can therefore offer better variations of physical properties than pure materials, such as graphene, since they can provide a larger number of compositional variables that can be tuned for specific properties. Figure 1 shows schematic side views of the $Ti_3AlC_2$, $Ti_3C_2T_x$, and TiC structures where the blue and the black spheres are the Ti and C atoms, respectively, with strong covalent bonds in the conductive carbide core layer. The stacking of the Ti and C atoms forms three monolayers of Ti and two monolayers of C in an alternated sequence. In the $Ti_3AlC_2$ the $Ti_3C_2$-layers are alternated with monolayers of Al, highlighted in Fig. 1 as yellow spheres, while the $Ti_3C_2T_x$ shows purple and red spheres on both sides of the transition metal carbide that are the F and O atoms, respectively, terminating the surfaces. Two alternatives for the F and O atoms to coordinate on the $Ti_3C_2$ surfaces are the *three-fold hollow face-centered cubic (fcc)* sites and the *three-fold hollow hexagonal close-packed (hcp)* sites [14-17]; an fcc site is formed by three surface Ti atoms in a triangular formation where the center of the triangle is above a Ti atom in the second (middle) Ti-monolayer while an hcp site is formed by three surface Ti atoms in a triangular formation where the center of the triangle is above a C atom in the adjacent C-monolayer. Other alternative coordination sites are on top of the surface Ti atoms or in *bridge sites* between two Ti atoms [14]. The fcc site (also called A-site) as the preferred site for the termination species on $Ti_3C_2T_x$ has been confirmed experimentally using high-resolution transmission electron microscopy (HRTEM) and X-ray photoelectron spectroscopy (XPS) [18]. The combined HRTEM/XPS study found that F only occupies the fcc sites in competition with O and that O also occupies other sites, *e.g.* bridge sites or on top sites, but not the hcp sites (also called B-sites). The study found no other terminating species than F and O and that F desorbs at elevated temperatures (>550 °C).

Despite vast interest in MXenes in general, and in $Ti_3C_2T_X$ in particular, there is little known experimentally about the bonding between the transition metals and the terminating species, $T_x$. Previous work of MXenes' electronic structure has mainly been based on ground-state density functional theory (DFT) calculations at 0 K [14-17,19]. Many of the theoretical investigations find no obstacles regarding replacing the inherently formed termination species with others in the pursuit of tailoring the properties for specific applications. Yet there are no indisputably experimental evidences showing that the inherently formed termination species can be replaced (re-termination) [20]. The different theoretical and experimental results and experiences show how important it is to fully understand the bonding conditions at the surfaces of the MXenes.

In this work, we will elucidate the local structural properties and interactions around the Ti atoms in $Ti_3AlC_2$ and $Ti_3C_2T_x$ through synchrotron radiation-based Ti *K*-edge (*1s*) X-ray absorption spectroscopy, including both X-ray absorption near-edge structure (XANES) and extended X-ray absorption fine structure (EXAFS). XANES probes the unoccupied density of states of the absorbing atoms and is therefore an ideal technique for determining the chemical surroundings and local bonding structure around the transition metal. The *1s* electron transition can, as a consequence of the selection rule, only occur to molecular orbitals with *p*-character (electric dipole transition) and *d*-character (electric quadrupole transition), although the probability of the *1s → 3d* transitions are about one thousandth of *1s → np* transitions [21,22]. EXAFS provides information about the coordination numbers, atomic distances, and amount of atomic displacements and disorder around the probed element [23].

Ti *1s* XANES and EXAFS spectra of $Ti_3AlC_2$ and $Ti_3C_2T_x$ have been presented previously [24-26]. Yet a detailed analysis of the spectra remains to be performed. Instead of only comparing the Ti *1s* XANES and EXAFS spectra of different compounds for similarities (or not) or a crude estimate of the Ti oxidation state, as in previous XANES and/or EXAFS studies of MAX-phases and MXenes [24-30], the present work aims to learn more about the local bonding around the probed Ti atoms in both $Ti_3AlC_2$ and $Ti_3C_2T_x$ through distinguished Ti *1s* XAS features.

The $Ti_3C_2T_x$ samples were fabricated as freestanding foils through wet etching, which leads to inherent F and O terminations [18]. Through heat treatments the fcc site coordinated F will desorb and be replaced by O, which prefer the fcc site when it is available [18]. The change of the fcc coordination will induce modifications in the XANES and EXAFS spectra and, thus, reveal local information of the bonding situation between the transition metal atoms and termination species. Hence, the study demonstrates how X-ray absorption spectroscopy can be used to probe the MXene surfaces to shed more light onto the local chemical interaction between F, O, and Ti atoms, which is relevant knowledge when tailor designing MXenes for after-sought material properties.

## 2. Experimental details

### 2.1 Sample preparation

Powders of $Ti_3AlC_2$ were produced starting with a mixture of TiC (Alfa Aesar, 98+%), Ti (Alfa Aesar, 98+%), and Al (Alfa Aesar, 98+%) of 1:1:2 molar ratios. The mixture was processed in a mortar with a pestle for 5 min and thereafter inserted in an alumina tube furnace. While a continuous stream of Ar gas, the furnace was heated with a rate of 5 °C min$^{-1}$ up to 1450 °C and held for 280 min before cooled down to room temperature. The resulting material is a lightly sintered $Ti_3AlC_2$ sample, which is then crushed to powder of particle size < 60 microns using mortar and pestle. A few mg of the $Ti_3AlC_2$ powder was mixed with polyethylene powder (Aldrich, 40-48 μm particle size) and thereafter cold pressed into a ~500 μm thick pellet for the X-ray absorption spectroscopy.

To convert the $Ti_3AlC_2$ to $Ti_3C_2T_x$ flakes, half a gram of $Ti_3AlC_2$ powder was added to a premixed 10 ml aqueous solution of 12 M HCl (Fisher, technical grade) and 2.3 M LiF (Alfa Aesar, 98+%) in a Teflon bottle. Prior to adding the MAX powder to the HCl/LiF(aq) solution, the latter was placed in an ice bath to avoid the initial overheating that otherwise can be a consequence of the exothermic reaction when the MAX power is added. After 0.5 h in the ice bath the bottle was placed on a magnetic stirrer hot plate in an oil bath and held at 35 °C for 24 h. The mixture was thereafter washed, first through 3 cycles using 40 ml of 1 M HCl(aq) and thereafter 3 cycles using 40 ml of 1 M LiCl (Alfa Aesar, 98+%). Then, the mixture was washed through cycles of 40 ml of deionized water until the supernatant reached a pH of approximately 6. In the end, 45 ml of deionized water were added, which was deaerated by bubbling $N_2$ gas through it and sonicated using an ultrasonic bath for 1 h. The resulting suspension was centrifuged for 1 h at 3500 rpm, which removed larger particles. The supernatant produced had a $Ti_3C_2T_x$ concentration of 1 mg ml$^{-1}$. To make freestanding films, 20 ml of the suspension were filtered through a nanopolypropylene membrane (3501 Coated PP, 0.064 μm pore size, Celgard, USA). The obtained $Ti_3C_2T_x$ foil is in detail described in Ref. [7].

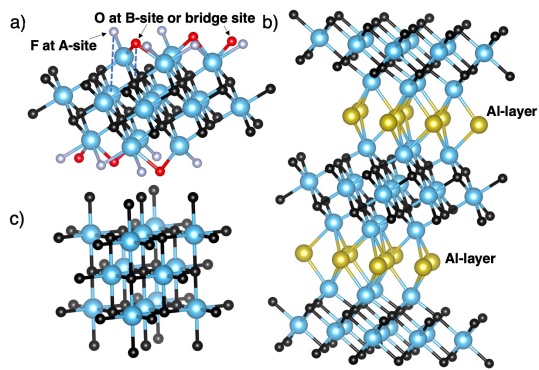

FIG. 1. Structure of an $M_3C_2T_x$ MXene layer with various termination sites of -$T_x$: -F (purple) and -O (red). The F atoms are adsorbed in a *three-fold hollow fcc site* (A-site). In this model structure, the O atoms are in a *bridge site* between two Ti and in a *three-fold hollow hcp site* (B-site). Blue and black spheres are Ti and C atoms, respectively, with strong covalent bonds in the $M_3X_2$ conductive carbide core layer.

Moreover, to observe XANES and EXAFS features originating from termination species, the surfaces must be free from oxidized material, mainly $TiO_2$ [31]. The new-made $Ti_3C_2T_x$ foils were therefore stored in argon (Ar) atmosphere and mounted on the sample holder in a glove-bag filled with nitrogen gas ($N_2$). A continuous flow of $N_2$ protected the sample from oxidation during measurement. Hence, the obtained XANES and EXAFS spectra of the $Ti_3C_2T_x$ foils have no detectable contribution from $TiO_2$ impurities. The insignificant amount of $TiO_2$ impurities and carbon containing contamination in the $Ti_3C_2T_x$ samples was confirmed through XPS. The XPS further showed a small amount of adsorbed Cl that desorbed completely at a moderate heat treatment. In addition, no indication of OH-termination was observed, which agrees with the combined HRTEM/XPS study [18]. The $Ti_3AlC_2$ (and TiC), on the other hand, showed small amounts of $TiO_2$ impurities in the XPS spectra. Contribution from $TiO_2$ impurities in the Ti *1s* XANES and EXAFS spectra of the $Ti_3AlC_2$ (and TiC) can therefore not be excluded. Nevertheless, the $TiO_2$ contribution showed to be at a negligible level.

## 2.2 XANES and EXAFS analysis

The XANES and EXAFS spectra were measured at the Ti *1s* edge using a Si111 crystal in the monochromator at the wiggler beamline BALDER on the 3 GeV electron storage ring at MAX IV in Lund, Sweden. The X-ray absorption of the $Ti_3C_2T_x$ MXene, the $Ti_3AlC_2$ MAX phase, a TiC reference, and a Ti metal reference foil were monitored in transmission mode with ionization chambers ($I_o$: 200 mbar $N_2$ and He; $I_1$: 2 bar $N_2$). The $Ti_3C_2T_x$ sample was positioned in a water-cooled gas cell (Linkam Scientific Instruments) with a low $N_2$ flow and a heating element that enabled measurements at different temperatures. The Ti *1s* XANES spectra were obtained at room temperature (RT) and at 250 °C after 20 minutes heat treatments at 250, 550, and 750 °C. The energy resolution at the Ti *1s* edge of the beamline monochromator was ~1.0 eV with 0.1 eV energy steps for XANES and 0.2 eV steps for EXAFS. The photon energy scale was calibrated using the first derivative of a Ti foil absorption spectrum, where the first inflection point was set to 4.9660 keV. The obtained spectra were normalized below the absorption edge before the background intensity subtraction and thereafter normalized above the absorption edge in the photon energy region of 5.045-5.145 keV. Self-absorption effects were found to be negligible in the normal incidence geometry for the $Ti_3C_2T_x$ and the Ti foils. The spectra of the $Ti_3AlC_2$ and TiC pellets showed, on the other hand, some self-absorption effects even at normal incidence and have therefore been corrected using the simple function $AI(\varepsilon)(1-BI(\varepsilon))^{-1}$, where A is a scaling factor, B is the self-absorption compensation factor, and $I(\varepsilon)$ is the Ti *1s* XANES spectrum of $Ti_3AlC_2$ and TiC, respectively. A and B are adjusted until the absorption features in the photon energy region of 5.045-5.145 keV have similar appearance as for the Ti *1s* XANES spectrum of $Ti_3C_2T_x$.

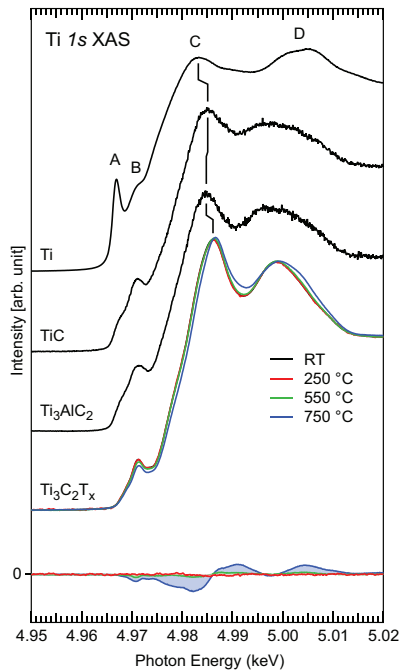

FIG. 2. Ti *1s* XANES spectra of Ti metal foil, TiC, Ti$_3$AlC$_2$, and Ti$_3$C$_2$T$_x$. The Ti$_3$C$_2$T$_x$ spectra are recorded at RT and at 250 °C after 20 minutes heat treatments at 250, 550, and 750 °C. Note that the red spectrum of Ti$_3$C$_2$T$_x$ heat-treated at 250 °C covers the black spectrum of non-heated Ti$_3$C$_2$T$_x$. The difference of the RT Ti$_3$C$_2$T$_x$ spectrum and the spectra obtained after 250, 550, and 750 °C heat treatment are shown at the bottom.

The Ti-Ti, Ti-C, Ti-O and Ti-F scattering paths obtained from the Effective Scattering Amplitudes (FEFF) [32-34] were included in the EXAFS fitting using the Visual Processing in EXAFS Researches (VIPER) software package [35]. The k$^2$-weighted χ EXAFS oscillations were extracted from the raw absorption data, the average of 10 absorption spectra, after removing known monochromator-induced glitches and subsequent atomic background subtraction and normalization. The atomic distances (R), number of neighbors (N), Debye-Waller factors ($\sigma^2$, representing the amount of disorder) and the reduced $\chi_r^2$ as the squared area of the residual, were determined by fitting the back-Fourier-transform signal between k=0–12.5 Å$^{-1}$ originally obtained from the forward Fourier-transform within R=0–3.35 Å of the first coordination shell using a Hanning window function [32-34] with a many body factor of $S_0^2$=0.8. The disorder and high-frequency thermal vibration of the atoms depending on the temperature was accounted for by an increasing Debye-Waller term $\sigma^2=\sigma^2_{stat}+\sigma^2_{vib}$ consisting of a static and a vibrational part that was proportional to the difference of the mean square atomic displacements [36-38].

## 3. Results and Discussion

Figure 2 shows the obtained high-resolution Ti *1s* XANES spectra of Ti-metal, TiC, Ti$_3$AlC$_2$, and Ti$_3$C$_2$T$_x$. The absorption spectra consist of two regions: pre-edge and main-edge. Features in the pre-edge region of *K*-edge XANES spectra of *3d* transition metal compounds are assigned to electric dipole (*1s* → *np*) and quadrupole (*1s* → *3d*) transitions. However, the intensity contribution of the latter is minor, because of orbital symmetry restrictions [21,22,39], and pre-edge features are therefore in most cases assigned to *1s* electron excitation into *p-d* hybridized orbitals, *i.e.* the transition into the *p*-component of the molecular orbitals that have both *p*- and *d*-character. The intensity of the pre-edge features depends mainly on coordination, symmetry, and bond angles [40].

The pre-edge features in the spectrum of the Ti metal reference, peaks A and B at 4.967 and 4.971 keV, respectively, are typical examples of *1s* → *p-d* hybridized molecular orbitals excitation near the Fermi energy (E$_f$) [41]. The sharp peak A near the Fermi energy originates from Ti *1s* excitation into a local Ti *3d* - *4p* mixing orbital (internal orbital mixing in the probed element), while peak B originates from Ti *1s* excitation into hybridized *3d* - *4p* orbital where the *3d* contribution originates from neighboring Ti atoms [42,43]. The intensity of the pre-edge features is, however, reduced because of the six-fold coordination of the Ti atoms in the metal [41]. The main-edge region consists of two broad features C (white line) and D that corresponds to *1s* → *4p* excitations [43]. With the Ti *1s* XANES spectrum of the Ti metal reference it is possible to estimate the E$_f$ position, which is located close to the on-set of peak A at 4.9642 keV. The full width at half maximum of peak A of the Ti metal reference (1.3 eV) is also an indication of the good resolution at the Balder beamline.

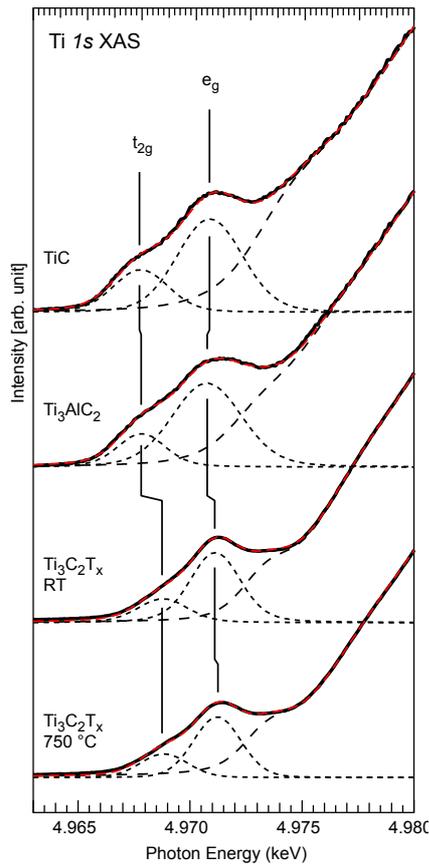

FIG. 3. Curve fitting of the Ti *1s* XANES pre-edge region for TiC, Ti$_3$AlC$_2$, room temperature (RT) Ti$_3$C$_2$T$_x$, and Ti$_3$C$_2$T$_x$ after 750 °C heat treatment. The dotted lines represent the Ti *1s* excitations into the $t_{2g}$ and $e_g$ orbitals and the dashed lines represents the Ti *1s* → *4p* excitations (black) and the accumulated intensity from all Ti *1s* excitations (red).

The XANES spectra of TiC, Ti$_3$AlC$_2$, and Ti$_3$C$_2$T$_x$ display significantly different structures compared with the Ti metal reference spectrum (see Fig. 2). The pre-edge region consists of two features, which is highlighted in Fig. 3. The TiC and Ti$_3$AlC$_2$ spectra show a shoulder at 4.9679 keV and a peak at 4.9710 and 4.9709 keV, respectively, while the Ti$_3$C$_2$T$_x$ spectrum shows a shoulder at 4.9689 keV and a peak at 4.9711 keV. A basic density of states calculation obtained from a simple Ti$_3$C$_2$T$_x$ model with T$_x$ being F on the fcc-sites and O on bridge sites shows two C *2p* peaks located at 4.2 and 7.1 eV above the E$_f$ and two Ti *3d* peaks located at 4.0 and 6.9 eV above E$_f$. The orbital mixing provides the C *2p* - Ti *3d* hybridization and would correspond to ~4.968 and ~4.971 keV, respectively, in the Ti *1s* XANES, which is close to the two pre-edge features obtained in the experimental spectra. Hence, the Ti *1s* XANES pre-edge features of TiC, Ti$_3$AlC$_2$, and Ti$_3$C$_2$T$_x$ are assigned to Ti *1s* → C *2p* - Ti *3d* hybridized molecular orbitals excitation. The Ti$_3$C$_2$T$_x$ spectrum shows lower intensity in the pre-edge region compared to TiC and Ti$_3$AlC$_2$, which suggests that the Ti *1s* → C *2p* - Ti *3d* hybridized molecular orbitals excitations are reduced for Ti$_3$C$_2$T$_x$ compared with both TiC and Ti$_3$AlC$_2$. The energy difference between the two C *2p* - Ti *3d* peaks corresponds the crystal-field splitting of the Ti *3d* states into the $t_{2g}$ and $e_g$ orbitals [43] and can, thus, experimentally be determined to be 3.1, 2.9, and 2.2 eV for TiC, Ti$_3$AlC$_2$, and Ti$_3$C$_2$T$_x$, respectively. The absence of a sharp peak near the E$_f$ in the Ti *1s* XANES of TiC, Ti$_3$AlC$_2$, and Ti$_3$C$_2$T$_x$ indicates that there is no local unoccupied Ti *3d-4p* hybridized orbital available for electron excitation.

The main-edge region shows two peaks, C and D in Fig. 2, which because of the 2D nature is relatively sharp for Ti$_3$C$_2$T$_x$. The peaks originate from Ti *1s* → *4p* excitations [43]. In addition, a closer look at the rising edge, *i.e.* the steep intensity increase at the main-edge, reveals a shoulder at 4.9780 keV. The shoulder is more noticeable in the Ti$_3$C$_2$T$_x$ spectrum, which is because of the high-energy shift of the first main-edge peak C; the peak C is at 4.9850 and 4.9847 keV for the TiC and Ti$_3$AlC$_2$ spectra, respectively, while the Ti$_3$C$_2$T$_x$ spectrum has the peak C at 4.9861 keV. The position of peak C for the TiC and Ti$_3$AlC$_2$ spectra is almost the same, which suggests that the interaction between the Al-layers and the Ti$_3$C$_2$-layers in Ti$_3$AlC$_2$ is very weak. The 1.4 eV shift of the C peaks between the Ti$_3$AlC$_2$ and Ti$_3$C$_2$T$_x$ spectra is a consequence of the replacement of the weak interacting Al-layers with the stronger interacting termination species T$_x$, where the F and O atoms attract charge from the Ti atoms.

In Fig. 2 there are also Ti *1s* XANES spectra of the Ti$_3$C$_2$T$_x$ sample after it has been heated to 250, 550 and 750 °C, respectively. After each heat treatment (to 550 and 750 °C) the sample was brought back to 250 °C before XANES-spectrum recording. As expected there are no significant changes in the spectra after the heat treatments to 250 and 550 °C, because temperatures above 550 °C are needed to alter the termination of the Ti$_3$C$_2$T$_x$ surfaces [18]. Included in Fig. 2 are also difference spectra that highlight the influence from the heat treatments of the Ti$_3$C$_2$T$_x$ sample. The small deviations from the zero line are inconsiderable for the temperatures 250 and 550 °C, although there is an indication of that 550 °C is the temperature threshold onset when to introduce changes in the T$_x$ coordination, which then are reflected in the Ti *1s* XANES spectra. A heat treatment to 750 °C (and subsequent cooling to 250 °C) provides, on the other hand, a high-energy shift of the peaks in the main-edge region of 0.5 eV, which has the effect that the pre-edge region appears to show a slightly reduced intensity. The difference spectrum for the 750 °C spectrum shows the characteristic variations common for a main-edge energy shift and not an intensity redistribution. It is interesting to note that the energy positions of the features in the pre-edge region are almost not affected by the heat treatment, see Fig. 3. The $t_{2g}$ peak has the same position while the $e_g$ peak has shifted 0.1 eV and, thus, widen the crystal-field splitting slightly. That the pre-edge region, which mainly originates from Ti *1s* excitations into the C *2p* - Ti *3d* hybridized molecular orbitals in the Ti$_3$C$_2$ layer, is almost unaffected by the heat treatments is suported by the previous combined HRTEM/XPS study that found that while removing F from a Ti$_3$C$_2$T$_x$ sample through a heat treatment the C *1s* XPS carbide peak remained unaffected [18]. That the crystal-field splitting to some extent widens suggest a stronger interaction between the O and the fcc-site compared to F. The main-edge features that originate from the Ti *1s* → *4p* excitation show, on the other hand, higher sensitivity toward the fcc-site occupation.

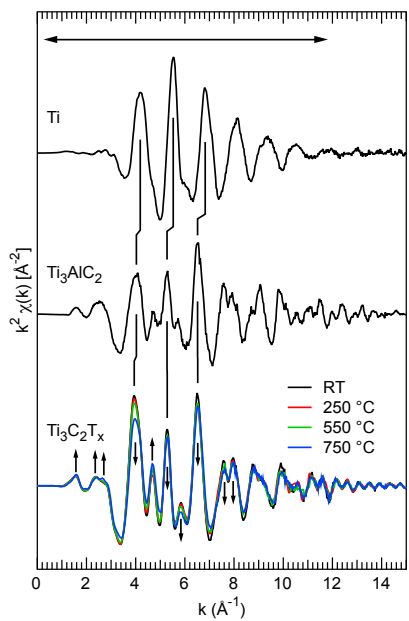

FIG. 4. $k^2$-weighted EXAFS, $k^2\chi(k)$, as a function of the photoelectron wave number k of the Ti metal foil, Ti$_3$AlC$_2$, and Ti$_3$C$_2$T$_x$ sample. The Ti$_3$C$_2$T$_x$ samples are recorded at RT and at 250 °C after 20 minutes heat treatments at 250, 550, and 750 °C. The horizontal arrow at the top shows the k-window for the most pronounced oscillations and the vertical arrows indicate changes in the peak intensities.

The XANES spectra of Ti$_3$C$_2$T$_x$ show some similarities with the XANES spectrum of TiO$_2$ [24-27,41,43,44]. A direct comparison shows that the pre-edge region of Ti$_3$C$_2$T$_x$ has more intensity and a smaller crystal-field splitting. The absorption rising edge of the Ti$_3$C$_2$T$_x$ spectrum is shifted about -3 eV while the absorption energy shift at peak C is about -1 eV. A larger difference between the Ti$_3$C$_2$T$_x$ and the TiO$_2$ XANES spectra is the shape of peak D where the TiO$_2$ shows strong peak intensity at 5.0035 keV that is absent in the Ti$_3$C$_2$T$_x$ spectrum before the heat treatment; hence, no detectable contribution from TiO$_2$ impurities in the Ti$_3$C$_2$T$_x$ sample. The Ti *1s* XANES spectra of the 750 °C heat treated Ti$_3$C$_2$T$_x$ show a trend of changes – slightly larger crystal-field splitting, 0.5 eV shift of the rising edge and the C peak, and the intensity shift in the D peak inducing a shoulder at 5.0035 keV – that suggest a stronger Ti-O interaction in heat treated Ti$_3$C$_2$T$_x$ compared to the non-treated Ti$_3$C$_2$T$_x$. Hence, the Ti *1s* XANES supports the observatiion [18] of that F occupies only the fcc-sites and that a heat treatment up to 750 °C removes F and make the fcc-sites available for O where the Ti-O interaction becomes stronger. (Experiments with and without the low N$_2$ flow in the Linkam gas cell ensured that no oxidation of the Ti$_3$C$_2$T$_x$ sample occurred in the presented work.)

Figure 4 shows the EXAFS structure factor oscillations of Ti$_3$AlC$_2$ and Ti$_3$C$_2$T$_x$ in comparison with the Ti metal reference, obtained from raw data that has not been phase shifted. The structure factors χ are displayed as a function of the wave vector k, that were $k^2$-weighted to highlight the higher k-region, where $k = \hbar^{-1}$ sqrt$[2m(E – E_o)]$ is the wave vector of the excited electron in the X-ray absorption process. The frequency of the oscillations and intensity of the EXAFS signal are directly related to the atomic distances (R) and the number of nearest neighbors (N), respectively; a higher frequency of the oscillations implies extended R while an enlarged amplitude implies increased N.

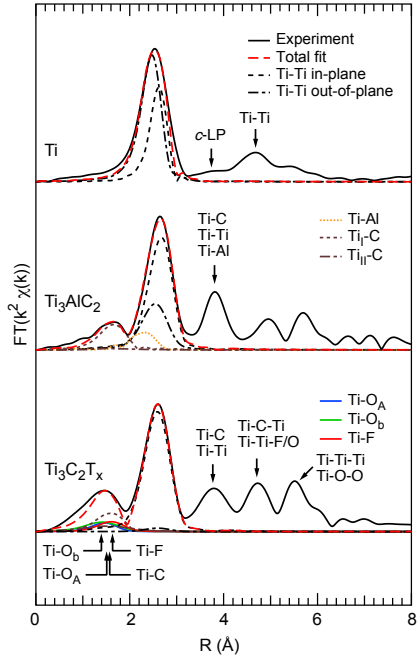

FIG. 5. Fourier transform obtained from the k²-weighted EXAFS oscillations χ(k) in Fig. 4 of Ti metal foil, $Ti_3AlC_2$, and $Ti_3C_2T_x$.

Starting with Ti metal at the top of Fig. 4, we observe the main oscillations at 4.18, 5.55, 6.80 Å$^{-1}$ where the middle one is caused by the Ti-Ti in-plane scattering that corresponds to the distance for the *a*-axis in the hexagonal crystal structure. For *k*-values above 12 Å$^{-1}$, the oscillations are damped out and simultaneously the noise increases.

For $Ti_3AlC_2$ and $Ti_3C_2T_x$, shown in the middle and bottom of Fig. 4, the main sharp oscillations occur in the 3.4-7.0 Å$^{-1}$ *k*-space region. There are also peaks at k=4.63 and 5.83 Å$^{-1}$, *i.e.* between the main Ti-Ti peaks, that only appear as weak shoulders in the Ti metal. The positions of the three main Ti-Ti peaks in $Ti_3AlC_2$ and $Ti_3C_2T_x$ (3.93, 5.28, 6.50 Å$^{-1}$) are shifted -0.25 to -0.30 Å$^{-1}$ in comparison to Ti metal (4.18, 5.55, 6.80 Å$^{-1}$) and similar as for TiC [45,46]. The small features at low-k values at 1.58 and 2.4-2.7 Å$^{-1}$ for the $Ti_3C_2T_x$ are associated with superimposed oscillations from Ti-O/F scattering.

While the heat treatments do not cause any *k*-shifts, the intensities of the oscillations decrease with increasing temperature, except for the oscillation at k=4.63 Å$^{-1}$. The oscillation intensity of the double peak at 7.55-7.95 Å$^{-1}$ also decreases with increasing temperature. More interestingly, the small features at 1.58 and 2.4-2.7 Å$^{-1}$ are also affected by the heat treatment. In order to analyze the detailed local structure and atomic distances in the films, Fourier transforms of the EXAFS data were performed.

Figure 5 shows the magnitude of the Fourier transform obtained from the k²-weighted EXAFS oscillations χ(k) in Fig. 4 by the standard EXAFS procedure [34] related to the radial distribution function. The horizontal arrow at the top of Fig. 4 indicates the applied *k*-window. Table I shows the final results of the EXAFS fitting using the FEFF scattering paths of $Ti_3AlC_2$ and $Ti_3C_2T_x$ and hcp Ti metal as structure model systems. The obtained radius values are in comparison to atomic distances determined for lattice parameters from X-ray diffraction (XRD) in the literature, listed in parenthesis in Table I. The initial crystal structure in the modelling assumes a $Ti_3C_2F_1O_2$ composition in line with previous quantitative core-level XPS results [18]. However, the obtained atomic distances have sources of errors such as photon energy calibration and dispersion, statistical noise, and inaccuracies in the *ab initio* calculations of scattering paths using FEFF. The corresponding errors in XRD are in the same order of magnitude as for EXAFS.

In Ti metal with hcp structure, the main peak consists of the in-plane Ti-Ti scattering (N=6) at 2.842 Å and (N=6) of 2.966 Å, where the latter corresponds to the *a*-axis of the crystal. For comparison, the peak that consists of the out-of-plane Ti-Ti scattering path of the *c*-axis is located at 4.675 Å as a weak feature. The corresponding values obtained from XRD measurements are 2.897, 2.950, and 4.686 Å, respectively [47]. The more intense peak in Ti metal observed at ~ 4.7 Å in Fig. 4 is because of the superposition of the three longer direct Ti-Ti scattering paths with atomic distances 5.079-5.110 Å in the higher order coordination shells.

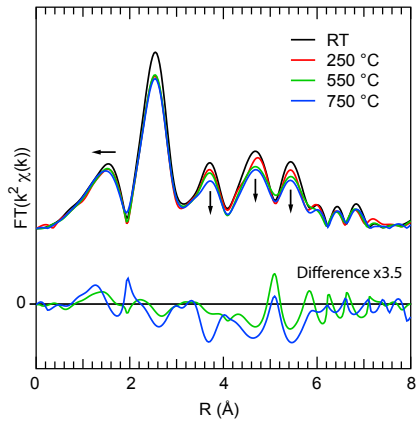

FIG. 6. Temperature-dependent Fourier transforms obtained from Ti *1s* EXAFS oscillations χ(k) of $Ti_3C_2T_x$. Difference spectra (x3.5) are shown at the bottom.

For $Ti_3AlC_2$, shown in the middle in Fig. 5, there is a slight shift of the main Ti-Ti peak to larger atomic distances and a new peak corresponding to Ti - C bonding appears at ~1.66 Å. In addition, there is a prominent peak at ~3.8 Å caused by Ti - Ti and Ti - Al scattering from the central Ti layer.

The main peak of the $Ti_3C_2T_x$, shown at the bottom in Fig. 5, is dominated (N=4.982) by the in-plane Ti-Ti scattering at 3.016 Å; the in-plane Ti-Ti scattering corresponds to the *a*-axis unit cell-edge. The weaker out-of-plane scattering at 3.072 Å shows a significantly lower intensity (N=1.312), which is not surprising for a 2D material. As also observed in Fig. 5 the Ti-Ti distances of the in-plane and out-of-plane contributions of $Ti_3C_2T_x$ (3.029 and 3.032 Å) are located at slightly different distances compared to the Ti-Ti atomic distances in $Ti_3AlC_2$ (3.040 and 3.111 Å). The value of 3.016 Å for the in-plane Ti-Ti distance of $Ti_3C_2T_x$ is somewhat smaller than the calculated in-plane Ti-Ti atomic distances obtained from the XRD lattice parameter (*a*=3.075 Å) [47].

The peak feature between 0 and 2 Å in the first coordination shell of $Ti_3C_2T_x$ is caused by superimposed Ti-O, Ti-F and Ti-C scattering where the Ti-C interaction has two different contributions; $Ti_I$ for the inner Ti atoms that bond only to C and a second $Ti_{II}$ contribution to the outer Ti atoms that bond both to C and the termination species. The Ti-C bond length of the outer Ti atoms is 0.056 Å longer than that for the inner Ti atoms. Interestingly, the three peaks in the higher coordination shells between 3-6 Å also contain a large contribution of *long inclined* single Ti-O and Ti-F scattering paths from all surface Ti atoms to the termination species $T_x$, in addition to Ti-Ti scattering and superimposed multi-scattering paths *e.g.*, Ti-Ti-C, Ti-C-O, Ti-C-F and Ti-F-Ti-F.

The $Ti-T_x$ scattering paths exhibit a significant temperature dependence as the data was measured at RT, 250, 550, and 750 °C. During heating, the peaks become significantly broadened and less intense as observed by the Debye-Waller factor that increases with the temperature, as a consequence of more atomic vibrations. For the main Ti-Ti scattering, $σ^2$ increases linearly from 0.0066 at room temperature to 0.0082 at 750 °C. A similar trend has been observed in other systems using EXAFS [48]. The broadening is also observed as increasing intensity between the peaks in the difference spectra. In addition to the broadening of the peaks, the vibrational behavior and peak broadening exhibit strong anisotropy of the Ti–Ti bonds. Therefore, before each measurement the elevated temperature was stabilized for 20 min and then decreased and stabilized at 250 °C after which the spectra were recorded.

Figure 6 shows the Fourier transform obtained from the $k^2$-weighted EXAFS oscillations χ(k) of $Ti_3C_2T_x$ and the effect of the heat treatment, where arrows indicate the general trends. Difference spectra that highlight the temperature-induced changes are shown in the bottom. In the low-radius region of the first coordination shell of the probed Ti (between 1-2 Å), including the bonding of the O and F elements, an intensity shift is observed corresponding to a ~0.2 Å shorter bond distance. In particular, the F-related intensity decreases indicating desorption of the F upon heating to 750 °C. Contrary to F, the intensity related to O-bonding slightly increases and shifts to shorter bond lengths. This behavior is consistent with previous temperature-dependent core-level XPS results [18] that showed desorption of F and a change of bond site for O in this temperature region.

**Table I:** Structural parameters for $Ti_3AlC_2$ and $Ti_3C_2T_x$ in comparison to Ti metal reference obtained from fitting of calculated scattering paths in the first coordination shell. N is the coordination number, R is the atomic distance (in Å) for the Ti-Ti and Ti-C scattering paths, respectively, σ is the corresponding Debye-Waller factor representing the amount of atomic displacement and disorder, reduced $\chi_r^2$ as the squared area of the residual, $N_{ind}$ is the number of independent points, P is the number of fitting parameters, and ν is the degrees of freedom. Atomic distances obtained from lattice parameters in XRD are given in parenthesis [39].

| System | Shell | R(Å)[a] | N[b] | σ(Å²)[c] | Statistics |
|---|---|---|---|---|---|
| Ti metal (at RT) | 6*Ti-Ti | 2.842 (2.897) | 5.535 | 0.0042 | $\chi^{0.95}{}_{20}$=8.44, $N_{ind}$=28, P=8, ν =N-P=20 |
| | 6*Ti-Ti: LP[d]-$a$ | 2.966 (2.950) | 2.237 | 0.0014 | |
| | 6*Ti-Ti: LP[d]-$c$ | 4.675 (4.686) | 2.005 | 0.0077 | |
| $Ti_3AlC_2$ (at RT) | 3*$Ti_{II}$-C | 2.160 (2.120) | 2.764 | 0.0025 | $\chi^{0.95}{}_4$=9.487, $N_{ind}$=24, P=20, ν=N-P=4 |
| | 6*$Ti_I$-C | 2.220 (2.227) | 1.560 | 0.0025 | |
| | 3*$Ti_{II}$-Al | 2.708 (2.781) | 1.299 | 0.0015 | |
| | 3*Ti-Ti oop[d] | 3.029 (3.069) | 1.164 | 0.0011 | |
| | 6*Ti-Ti: LP[d]-$a$ | 3.032 (3.075) | 5.735 | 0.0066 | |
| $Ti_3C_2T_x$ (at RT) | 1*$Ti_{II}$-$O_A$ | 1.707 | 0.199 | 0.0045 | $\chi^{0.95}{}_{10}$=8.307, $N_{ind}$=36, P=26, ν=N-P=10 |
| | 1*$Ti_{II}$-$O_b$ | 1.952 | 0.292 | 0.0030 | |
| | 1*$Ti_{II}$-F | 2.115 | 0.245 | 0.0029 | |
| | 6*$Ti_I$-C | 2.139 | 1.992 | 0.0012 | |
| | 3*$Ti_{II}$-C | 2.234 | 2.057 | 0.0079 | |
| | 6*Ti-Ti: LP[d]-$a$ | 3.016 (3.075) | 4.982 | 0.0056 | |
| | 3*Ti-Ti oop[e] | 3.072 (3.135) | 1.312 | 0.0021 | |

[a] the error in the atomic distances are estimated to be ±0.01 Å.
[b] the error in the coordination numbers are estimated to be ±0.01.
[c] the error in the Debye-Waller factors are estimated to be ±0.001 Å².
[d] Lattice parameter.
[e] out-off plane.

The three peaks observed between 3-6 Å in $Ti_3C_2T_x$ also occur in EXAFS data of cubic TiC [45,46]. The first peak at ~3.5 Å is mainly ascribed to Ti-Ti and Ti-C scattering in the second coordination shell. The second peak at 4.5-5 Å mainly consists of Ti-C-Ti and Ti-Ti-O/F scattering while the third peak 5.5-6 Å contain many multi-scattering paths such as Ti-Ti-Ti (5.30 Å) and Ti-Ti-Ti-C (5.31 Å) etc. However, the intensity of the three peaks in the higher coordination shells also contains a significant contribution of *long inclined* single Ti-O (3.62, 3.73, 4.28, 4.75, 4.79, and 5.04 Å) and Ti-F (3.76, 4.58 Å, 4.78, and 4.86 Å) scattering paths. As observed in the difference spectra, the contribution of these paths decrease as the temperature is increased. This is consistent with the fitting results showing that the Ti-O bond length in the first coordination shell seems to shorten as the temperature increases to 750 °C.

Yet, after a decade of extensive research activities, there is new knowledge to gain about MAX phases and MXenes. In the present study there are several interesting observations. For example, through XANES we find that the C *2p* - Ti *3d* hybridization is altered when the Ti$_3$AlC$_2$ transforms into Ti$_3$C$_2$T$_x$ leading to a smaller crystal-field splitting of the $t_{2g}$ and $e_g$ orbitals, which suggests slightly weaker Ti-C bonds in Ti$_3$C$_2$T$_x$ compared to Ti$_3$AlC$_2$. Another observation is the 1.4 eV energy shift of the main absorption edge for the Ti$_3$C$_2$T$_x$ compared to Ti$_3$AlC$_2$. A main-edge shift is often an indication of a charge redistribution and when the shift is toward higher energies the charge transfer is away from the probed atoms. Hence, replacing the weak interacting Al-layers in Ti$_3$AlC$_2$ with chemisorbed F and O in Ti$_3$C$_2$T$_x$ will withdraw charge from the Ti toward the termination species.

The additional main-edge energy shift caused by the heat treatment can, however, not be a consequence from a further withdrawal of charge from the Ti atoms, because that would contradict the previous temperature-programmed XPS study [18]; the intensity at the high binding energy side of the Ti *2p$_{3/2}$* XPS spectra decreases while F desorbs indicating that Ti in Ti$_3$C$_2$T$_x$ chemically reduces in a heat treatment. In addition, the electronegativity of O is lower compared to F (Pauling scale 3.44 and 3.98, respectively). The Ti *1s* XANES main-edge shift toward higher energy must therefore be caused by something else. It can, for example, be a response to the stronger bonding of the O in the fcc site compared to F that pushes the unoccupied Ti *4p* orbitals toward higher energy. Regardless of the reason, the Ti *1s* XANES shows that orbitals with Ti *4p* character are sensitive to changes of the termination species on the fcc-sites.

From the EXAFS we learn that the in-plane Ti-Ti distances decreases while the out-of-plane Ti-Ti distances increases when the Ti$_3$AlC$_2$ is converted into Ti$_3$C$_2$T$_x$. Concerning EXAFS, it probes the local short-order atomic distances between the absorber atom and the neighboring scatterers using the constructive and destructive interference in the unoccupied electronic structure. Since the EXAFS photoelectrons travel much faster than the speed of the thermal motion of the atoms the obtained atom distances are an average of "snapshots" that in most cases correspond to the distance between the average atomic positions as obtained with XRD and neutron diffraction, which are considered to be long-order probes. However, if adjacent atoms are moving in an anti-correlated motion the distance between them will be the same as the atomic positions distance, only when both atoms are in-plane, while it will become larger when the atoms are moving in opposite direction out-of-plane. The atom distances obtained from EXAFS will then be larger than if they would be obtained from XRD and neutron diffraction. The surface Ti atoms and the termination species F and O are expected to move in an anti-correlated motion and the obtained Ti$_{II}$-O$_A$, Ti$_{II}$-O$_b$, and Ti$_{II}$-F distances are probably larger than the true atomic positions. This discrepancy between the average atom distances between the T$_x$ atoms and the surface Ti atoms will become larger with increasing temperature.

## 4. Conclusions
Through a combination of XANES and EXAFS we have investigated the MAX phase material Ti$_3$AlC$_2$ and the MXene material Ti$_3$C$_2$T$_x$, where the latter was examined before and after a series of heat treatments. The pre-edge absorption region of both Ti$_3$AlC$_2$ and Ti$_3$C$_2$T$_x$ shows mainly Ti *1s* excitations into two C *2p* - Ti *3d* hybridized molecular orbitals corresponding to the Ti *3d* $t_{2g}$ and $e_g$ orbitals. The crystal-field splitting is determined to be 2.9 and 2.2 eV for the Ti$_3$AlC$_2$ and Ti$_3$C$_2$T$_x$, respectively. The main-edge absorption features originate from the Ti *1s → 4p* excitation and appears to be sensitive towards the fcc-site occupation, which lead to a 1.4 eV shift when the Al-layers in Ti$_3$AlC$_2$ were replaced with the termination species F and O. The local chemical bonding structure and structural properties with atomic distances in Ti$_3$C$_2$T$_x$ MXene shows significant temperature-dependence. Heat treatment up to 750 °C removed F and made the fcc-sites available for O occupation, which is manifested as a 0.5 eV high-energy shift of the peaks in the main-edge absorption region.

EXAFS shows that the shortest in-plane Ti-Ti atomic distances in Ti$_3$AlC$_2$ and Ti$_3$C$_2$T$_x$ is 3.032 and 3.016 Å, respectively, which are longer and shorter than the out-of-plane distance of 3.029 Å in Ti$_3$AlC$_2$ and the corresponding atomic distance in Ti$_3$C$_2$T$_x$ is 3.072 Å. The Ti$_I$-C and Ti$_{III}$-C bond lengths in Ti$_3$AlC$_2$ are 2.220 and 2.160 Å, respectively, while the Ti$_I$-C and Ti$_{II}$-C bond lengths in Ti$_3$C$_2$T$_x$ are 2.139 and 2.234 Å, respectively. Significant changes in the Ti-O/F coordination are observed with increasing temperature in the heat treatment. The Ti$_{III}$-O bond lengths becomes shorter because of a change in coordination from bridge to fcc facilitated through the desorption of the F as the F contribution is found to diminish when the temperature is raised from room temperature up to 750 °C. Significant contribution of *long inclined* single Ti-O and Ti-F scattering paths decrease as the temperature is increases.

## Acknowledgements

We acknowledge MAX IV Laboratory for time on Beamline Balder under Proposals 20190399 and 20191016. Research conducted at MAX IV, a Swedish national user facility, is supported by the Swedish Research council under contract 2018-07152, the Swedish Governmental Agency for Innovation Systems under contract 2018-04969, and Formas under contract 2019-02496. The computations were enabled by resources provided by the Swedish National Infrastructure for Computing (SNIC) at the National Supercomputer Centre (NSC) partially funded by the Swedish Research Council through grant agreement no. 2016-07213. We would also like to thank the Swedish Research Council (VR) LiLi-NFM Linnaeus Environment and Project Grant No. 621-2009-5258. The research leading to these results has received funding from the Swedish Government Strategic Research Area in Materials Science on Functional Materials at Linköping University (Faculty Grant SFO-Mat-LiU No. 2009-00971). M.M. acknowledges financial support from the Swedish Energy Research (Grant No. 43606-1) and the Carl Tryggers Foundation (CTS16:303, CTS14:310). Most importantly, we thank Dr. Joseph Halim at Linköping University for preparing the samples and Kajsa Sigfridsson Clauss at the MAX IV Laboratory for experimental support.

## References

[1] S. Novoselov, A. K. Geim, S. V. Morozov, D. Jiang, Y. Zhang, S. V. Dubonos, I. V. Grigorieva, and A. A. Firsov, *Electric Field Effect in Atomically Thin Carbon Films*, Science **306**, 666 (2004).

[2] M. Naguib, M. Kurtoglu, V. Presser, J. Lu, J. Niu, M. Heon, L. Hultman, Y. Gogotsi, and M. W. Barsoum, *Two-Dimensional Nanocrystals Produced by Exfoliation of Ti$_3$AlC$_2$*, Adv. Mater. **23**, 4248 (2011).

[3] M. Naguib, O. Mashtalir, J. Carle, V. Presser, J. Lu, L. Hultman, Y. Gogotsi, and M. W. Barsoum, *Two-Dimensional Transition Metal Carbides*, ACS Nano **6**, 1322 (2012).

[4] M. Naguib, V. N. Mochalin, M. W. Barsoum, and Y. Gogotsi, *25th Anniversary Article: MXenes: A New Family of Two-Dimensional Materials*, Adv. Mater. **26**, 992 (2014).

[5] Z. Yan, J. Yao, Z. Sun, Y. Zhu, and J. M. Tour, *Controlled Ambipolar-to-Unipolar Conversion in Graphene Field-Effect Transistors Through Surface Coating with Poly(ethylene imine)/Poly(ethylene glycol) Films*, Small **8**, 59 (2012).

[6] Y. Zhou, C. Fuentes-Hernandez, J. Shim, J. Meyer, A. J. Giordano, H. Li, P. Winget, T. Papadopoulos, H. Cheun, and J. Kim, *A Universal Method to Produce Low–Work Function Electrodes for Organic Electronics*, Science **336**, 327 (2012).

[7] M. Ghidiu, M. R. Lukatskaya, M.-Q. Zhao, Y. Gogotsi, and M. W. Barsoum, *Conductive two-dimensional titanium carbide 'clay' with high volumetric capacitance*, Nature **516**, 78 (2014).

[8] M. Naguib, J. Halim, J. Lu, K. M. Cook, L. Hultman, Y. Gogotsi, and M. W. Barsoum, *New Two-Dimensional Niobium and Vanadium Carbides as Promising Materials for Li-ion Batteries*, J. Am. Chem. Soc. **135**, 15966 (2013).

[9] N. Li, X. Chen, W.-J. Ong, D. R. MacFarlane, X. Zaho, A. K. Cheetham, and C. Sun, *Understanding of Electrochemical Mechanisms for $CO_2$ Capture and Conversion into Hydrocarbon Fuels in Transition-Metal Carbides (MXenes)*, ACS Nano **11**, 10825 (2017).

[10] J. Halim, M. R. Lukatskaya, K. M. Cook, J. Lu, C. R. Smith, L.-Å. Näslund, S. J. May, L. Hultman, Y. Gogotsi, P. Eklund, and M. W. Barsoum, *Transparent Conductive Two-Dimensional Titanium Carbide Epitaxial Thin Films*, Chem. Mater. **26**, 2374 (2014).

[11] X. Li, X. Yin, C. Song, M. Han, H. Xu, W. Duan, L. Cheng, and L. Zhang, *Self-Assembly Core–Shell Graphene-Bridged Hollow MXenes Spheres 3D Foam with Ultrahigh Specific EM Absorption Performance*, Adv. Funct. Mater. **28**, 1803938 (2018).

[12] M. W. Barsoum, *MAX Phases: Properties of Machinable Ternary Carbides and Nitrides* (Wiley, New York, 2013).

[13] M. Sokol, V. Natu, S. Kota, and M. W. Barsoum, *On the chemical diversity of the MAX phases*, Trends in Chemistry **1**, 210 (2019).

[14] M. Khazaei, M. Arai, T. Sasaki, C.-Y. Chung, N. S. Venkataramanan, M. Estili, Y. Sakka, and Y. Kawazoe, *Novel Electronic and Magnetic Properties of Two-Dimensional Transition Metal Carbides and Nitrides*, Adv. Func. Mat. **23**, 2185 (2013).

[15] Y. Xie and P. R. C. Kent, *Hybrid density functional study of structural and electronic properties of functionalized $Ti_{n+1}X_n$ (X=C, N) monolayers*, Phys. Rev. B **87**, 235441 (2013).

[16] Q. Tang, Z. Zhou, and P. Shen, *Are MXenes promising anode materials for Li ion batteries? Computational studies on electronic properties and Li storage capability of $Ti_3C_2$ and $Ti_3C_2X_2$ (X = F, OH) monolayer*, J. Am. Chem. Soc. **134**, 16909 (2012).

[17] Z. H. Fu, Q. F. Zhang, D. Legut, C. Si, T. C. Germann, T. Lookman, S. Y. Du, J. S. Francisco, and R. F. Zhang, *Stabilization and strengthening effects of functional groups in two-dimensional titanium carbide*, Phys. Rev. B **94**, 104103, (2016).

[18] I. Persson, L.-Å. Näslund, J. Halim, M. W. Barsoum, V. Darakchieva, J. Palisaitis, J. Rosen, and P. O. Å. Persson, *On the organization and thermal behavior of surface functional groups on $Ti_3C_2$ MXene in vacuum*, 2D Mater. **5**, 015002 (2017).

[19] T. Hu, M. Hu, B. Gao, W. Li, and X. Wang, *Screening Surface Structure of MXenes by High-Throughput Computation and Vibrational Spectroscopic Confirmation*, J. Phys. Chem. C **122**, 18501 (2018).

[20] P. O. Å. Persson and J. Rosen, *Current state of the art on tailoring the MXene composition, structure, and surface chemistry*, Current Opinion Solid State Mater. Sci. **23**, 100774 (2019).

[21] R. A. Bair and W. A. Goddard III, *Ab initio studies of the x-ray absorption edge in copper complexes. I. Atomic $Cu^{2+}$ and $Cu(II)Cl_2$*, Phys. Rev. B **22**, 2767 (1980).

[22] J. Kawai, *Absorption techniques in X-ray spectroscopy*, In Encyclopedia of Analytical Chemistry, Meyers RA (ed.). (Wiley: Chichester, 2000; 13288).

[23] v. S. R. Bordwehr, *A History of X-ray absorption fine structure*, Am. Phys. Fr. **14**, 377 (1989).

[24] M. R. Lukatskaya, S.-M. Bak, X. Yu, X.-Q. Yang, M. W. Barsoum, and Y. Gogotsi, *Probing the Mechanism of High Capacitance in 2D Titanium Carbide Using In Situ X-Ray Absorption Spectroscopy* Adv. Energy Mater. **5**, 1500589 (2015).

[25] Z. Li, L. Yu, C. Milligan, T. Ma, L. Zhou, Y. Cui, Z. Qi, N. Libretto, B. Xu, J. Luo, E. Shi, Z. Wu, H. Xin, W. N. Delgass, J. T. Miller, and Y. Wu, *Two-dimensional transition metal carbides as supports for tuning the chemistry of catalytic nanoparticles*, Nat. Commun. **9**, 5258 (2018).

[26] D. Zhao, Z. Chen, W. Yang, S. Liu, X. Zhang, Y. Yu, W.-C. Cheong, L. Zheng, F. Ren, G. Ying, X. Cao, D. Wang, Q. Peng, G. Wang, and C. Chen, *MXene ($Ti_3C_2$) Vacancy-Confined Single-Atom Catalyst for Efficient Functionalization of $CO_2$*, J. Am. Chem. Soc. **141**, 4086 (2019).

[27] E. Piskorska, K. Lawniczak-Jablonska, I.N. Demchenko, R. Minikayev, E. Benko, P. Klimczyk, A. Witkowska, E. Welter, and M. Heinonen, *Characterization of the c-BN/TiC, $Ti_3SiC_2$ systems by element selective spectroscopy*, J. Alloys Compd **382**, 187 (2004).

[28] E. Benko, P. Klimczyk, S. Mackiewicz, T.L. Barr, and E. Piskorska, *cBN–$Ti_3SiC_2$ composites*, Diam. Relat. Mater. **13**, 521 (2004).

[29] Z. Li, Y. Cui, Z. Wu, C. Milligan, L. Zhou, G. Mitchell, B. Xu, E. Shi, J. T. Miller, F. H. Ribeiro, and Y. Wu, *Reactive metal–support interactions at moderate temperature in two-dimensional niobium-carbidesupported platinum catalysts*, Nat. Catal. **1**, 349 (2018).

[30] E. B. Deeva, A. Kurlov, P. M. Abdala, D. Lebedev, S. M. Kim, C. P. Gordon, A. Tsoukalou, A. Fedorov, and C. R. Müller, *In Situ XANES/XRD Study of the Structural Stability of Two-Dimensional Molybdenum Carbide $Mo_2CT_x$: Implications for the Catalytic Activity in the Water−Gas Shift Reaction*, Chem. Mater. **31**, 4505 (2019).

[31] M. Magnuson, J. Halim, and L.-Å. Näslund, *Chemical Bonding in Carbide MXene Nanosheets*, J. Elec. Spec. **224**, 27 (2018).

[32] J. J. Rehr, J. J. Kas, F. D. Vila, M. P. Prange, and K. Jorissen, *Parameter-free calculations of X-ray spectra with FEFF9*, Phys. Chem. Chem. Phys. **12**, 5503 (2010).

[33] J. J. Rehr, J. J. Kas, M. P. Prange, A. P. Sorini, Y. Takimoto, and F. D. Vila, *Ab initio theory and calculations of X-ray spectra*, C. R. Physique **10**, 548 (2009).

[34] J. J. Rehr and R. C. Albers, *Theoretical Approaches to X-Ray Absorption Fine Structure,* Rev. Mod. Phys. **72**, 621 (2000).

[35] K. V. Klementev, *Extraction of the fine structure from X-ray absorption spectra*, J. Phys. D: Appl. Phys. **34**, 209 (2001).

[36] E. A. Stern, *Theory of the extended x-ray-absorption fine structure*, Phys. Rev. B **10**, 3027 (1974).

[37] E. A. Stern, *Structural determination by X-ray Absorption*, Contemp. Phys. **19**, 239 (1978).

[38] E. A. Stern and S. M. Heald, *Basic principles and applications of EXAFS; Handbook of synchrotron radiation*, (New York: North-Holland, 1983).

[39] T. Yamamoto, *Assignment of pre-edge peaks in K-edge X-ray absorption spectra of 3d transition metal compounds: electric dipole or quadrupole?*, X-Ray Spectrom. **37**, 572 (2008).

[40] N. Jiang, D. Su, and J. C. H. Spence, *Determination of Ti coordination from pre-edge peaks in Ti K-edge XANES*, Phys. Rev. B **76**, 214117 (2007).

[41] F. Farges, G. E. Brown Jr., and J. J. Rehr, *Ti K-edge XANES studies of Ti coordination and disorder in oxide compounds: Comparison between theory and experiment*, Phys. Rev. B **56**, 1809 (1997).

[42] D. Cabaret, A. Bordage, A. Juhin, M. Arfaouia, and E. Gaudryad, *First-principles calculations of X-ray absorption spectra at the K-edge of 3d transition metals: an electronic structure analysis of the pre-edge*, Phys. Chem. Chem. Phys. **12**, 5619 (2010).

[43] C. A. Triana, C. Moyses Araujo, R. Ahuja, G. A. Niklasson, and T. Edvinsson, *Electronic transitions induced by short-range structural order in amorphous $TiO_2$*, Phys. Rev. B **94**, 165129 (2016).

[44] T. C. Rossi, D. Grolimund, M. Nachtegaal, O. Cannelli, G. F. Mancini, C. Bacellar, D. Kinschel, J. R. Rouxel, N. Ohannessian, D. Pergolesi, T. Lippert, and M. Chergui, *X-ray absorption linear dichroism at the Ti K edge of anatase $TiO_2$ single crystals*, Phys. Rev. B 100, 245207 (2019).

[45] A. Balzarotti, M. De Crescenzi, and L. Incoccia, *Electronic relaxation effects on x-ray spectra of titanium and transition-metal carbides and nitrides*, Phys. Rev. B **25**, 6349 (1982).

[46] C. Adelhelm, M. Balden, and M. Sikora, *EXAFS investigation of the thermally induced structuring of titanium-doped amorphous carbon films*, Materials Sci. Eng. C **27**, 1423 (2007).

[47] "JCPDS - International Centre for Diffraction Data, Titanium metal (Ti): Ref ID 00-044-1294.".

[48] W. S. Chu, S. Zhang, M. J. Yu, L. R. Zheng, T. D. Hu, H. F. Zhao, A. Marcelli, A. Bianconi, N. L. Saini, W. H. Liu, and Z. Y. Wu, *Correlation between local vibrations and metal mass in $AlB_2$-type transition-metal diborides*, Journal Synchr. Rad. **16**, 30 (2009).